\documentclass[reprint,amsmath,amssymb,aps,prb]{revtex4-1}
\usepackage{graphicx}
\usepackage{dcolumn}
\usepackage{bm}

\begin{document}


\title{Phenomenological studies of femtosecond laser ablation\\ on optical thin films for integrated photonics}


\author{R Peyton}
	\email[]{robertop@ciop.unlp.edu.ar} 
\author{V Guarepi}%
\author{G A Torchia}%
	\affiliation{Centro de Investigaciones \'Opticas (CONICET-CICBA-UNLP), M.B. Gonnet (1897), Buenos Aires, Argentina}


\date{\today}

\begin{abstract}
In this work, for the first time of our knowledge, we present a well-supported procedure to fabricate ridge optical waveguides onto thin films of PLZT by femtosecond laser ablation. In order to achieve smooth guiding structures to guarantee good optical performance, we have studied the ablation threshold characteristics for different kinematic conditions of fs laser machining and also we have explored the different ablation regimens for several fluencies reached. Besides, we characterize the morphology and roughness of ridge waveguides through a modal and scattering loss analysis, for waveguides made with single and multiple scan. An innovative phenomenological model that relates the sidewall roughness to process parameters is proposed in this paper. Conducting this approach it is possible to extend the fabrication method of smooth guiding structures by fs micromachining to many optical thin films in order to develop integrated photonics devices addressed to different technological applications.
\end{abstract}


\maketitle

\section{Introduction}

The integrated photonic devices are currently used in many technological applications, including optical communications systems, microfluidics and sensors\cite{Lifante_2003,Fainman_2010,Agrell_2016}. According to the selected application, the materials and manufacturing technologies have to be appropriately chosen. For instance, electro-optical materials are very attractive in order to fabricate intensity optical modulators, nevertheless, they hardly ever must be used in mass-produced process according to the currently data transmission demand driven by internet\cite{Liu_2015}. 

An interesting material to be used in integrated photonics is lanthanum-modified lead zirconate titanate, well-known as PLZT (Pb,La)(Zr,Ti)O$_3$\cite{Haertling_1987}. This transparent ferroelectric ceramic has been recognized as a low-cost material for fast electro-optic devices, with high transparency at visible and infrared frequencies\cite{kawaguchi_1984}. Also, ultrafast nonlinear response at 100 GHz in a coplanar waveguide structure has been shown\cite{Bao_1995}. PLZT has presented an electro-optic coefficient at 612 pm/V (almost twenty times higher than LiNb03), which is one of the most interesting attribute\cite{nashimoto_2001}. However, this material is not compatible with many standard fabrication techniques. Consequently, the ultrafast laser micromachining can be used for that\cite{Zhang_2016}.

The femtosecond laser micromachining is a procedure to remove or change a material’s properties, which has been used in a large range of applications, from optical waveguide fabrication to cell ablation. This method usually has been performed on many field of applications as it is reported in \cite{Osellame_2012}. It presents unique capabilities which stand out from other manufacturing techniques. The most important features are the three dimensional processing, the rapid prototyping and the variety of materials that can be employed, including glasses, crystals, ceramics and polymers\cite{Gattass_2008,Phillips_2015}. Besides, fs-micromachining is an  economical process compared to conventional lithographic techniques, which needs expensive and complex environment to be implemented. This ductile method also can be combined with other technologies. For those reasons, the fs-micromachining is considered an easily and directly technique of material processing. In particular in this work, we paid attention in ridge optical waveguides constructed through the ultrafast ablation mechanism over thin film. The procedure consists of micromachining two parallel ablation grooves along the surface sample that keeps the light confined transversally thus shaping the waveguide mode\cite{Chen_2014}. This offers an alternative pathway to construct ridge waveguides on planar waveguide substrates, which could be produced by any other standard techniques. 

The ridge waveguides are particularly used in the fabrication of microfluidic chips, sensors and active photonics devices; in this sense many works have been reported\cite{Chen_2014,Degl_2006,Sun_2007,dong_2003}. However, an important drawback of these guiding structures is the high rough sidewalls produced by the laser ablation, so this effect introduces considerable losses and degrades the waveguide quality\cite{Chen_2014}. Another unfortunate effect is the deposition of the debris ablated material onto surface of the sample, increasing thus, the scattering losses\cite{Gottmann_2007}. Generally, in order to reduce the roughness of the waveguides, post-ablation-treatment (such as ion-beam sputtering)\cite{Degl_2006}, thermal annealing or multiple scans could be performed\cite{Sun_2007}. Although there are frameworks that describe the process of ablation from the atomic and structural point of view, but they do not relate the roughness of the ablated grooves\cite{Bouilly_2007,Christensen_2009}. Therefore, one of the most important goals would be the relationship between the process parameters and the ablation morphology and the source of the sidewall roughness, as well also we need to know this dependence on the materials or samples studied. 

In this article, we fabricate ridge optical waveguides on thin PLZT films by femtosecond laser ablation technique. In particular, we study the ablation threshold for different kinematic conditions and the ablation regimens for several fluencies, in order to achieve smooth structures. Additionally, we propose a phenomenological model that relates the sidewall roughness to the process parameters. An ablation quality factor that takes into account the process randomness has been introduced. Finally, we characterize the morphology and roughness of ridge waveguides through the modal analysis and scattering losses, respectively, using one and multiple writing scans during the fabrication process.

\section{Materials and procedures}

The ridge optical waveguides were fabricated using a commercial Ti:sapphire ultrafast laser system (Spectra Physics Spitfire) working at 793 nm. Pulse duration was 173~fs at 1~kHz repetition rate. The pulse energy was adjusted combining a half-wave plate plus a Glan polarizer and a neutral density filter of 1.3~OD. A 10x (NA~=~0.25) microscope objective was used to focus the laser on the sample surface and the spot size was $w_0$~=~2.9 $\mu$m at 1/e$^2$. A micrometer motorized station system (Newport, Inc) with an accuracy at about $\pm$275 nm and a shutter with a time resolution of 1 ms were used. Furthermore, the plasma emission during the laser ablation was captured by a HR2000+ spectrometer (OceanOptics, Inc.).

Pb$_{0.91}$La$_{0.09}$Zr$_{0.65}$Ti$_{0.35}$O$_3$ thin films with a thickness of around 500~nm were prepared by the chemical solution deposition technique on SiO$_2$/Si substrates. The amorphous films were deposited by spin coating technique (CSD) and thermal treated by Rapid Thermal Processing for film crystallization. The PLZT films have exhibited a well-crystallized perovskite structure\cite{Pellegri_2000}. 

The morphology and roughness were analyzed using a scanning electron microscope system (Quanta~200, Thermo Fisher Scientific). The end-coupling technique was used to characterize the ridge optical waveguides\cite{Tong_2014}. The spatial intensity distribution of modes was measured with a Thorlabs BC106N-VIS beam profiler analyzer by placing a microscope objective at the output of the waveguides. The light source used in the experiment was a 532~nm laser diode with 50~mW optical power. All experiments were carried out in open atmosphere conditions.

\section{Results and discussions}

\subsection{Ablation threshold}

\begin{figure}[t]
\includegraphics[width=0.45\textwidth]{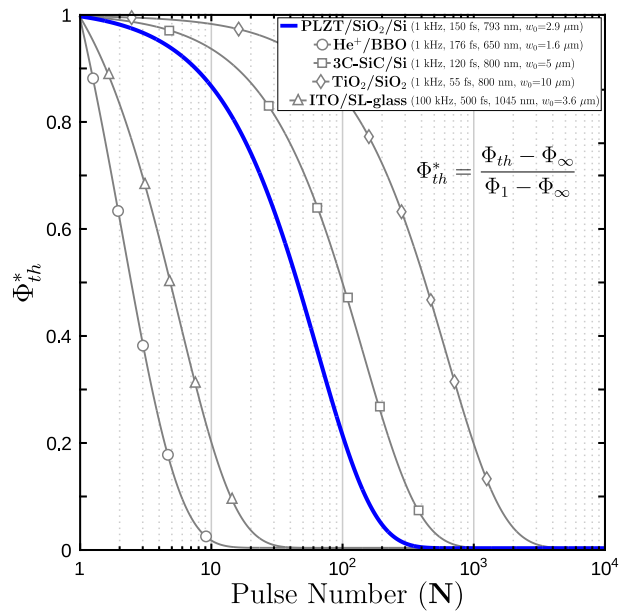}
\caption{\label{fig:01} Non-dimensional ablation threshold fluences as a function of the number of incident pulses for different types of thin film planar waveguides\cite{Degl_2006,sun_2015,dong_2003,choi_2007}. The experimental values obtained for PLZT thin films are $\Phi_{1}$~=~6.1~J/cm$^{2}$, $\Phi_{\infty}$~=~2.9~J/cm$^{2}$ and $k$~=~0.015.}
\end{figure}

In order to choose correctly the machining parameters, it is necessary consider the ablation characteristics of the samples. In this sense, the ablation threshold fluence $\Phi_{th}$ is the most important feature. This parameter is related to the number of laser shots $\mathbf{N}$ and an incubation factor $k$, which depend on the micromachining kinematic and the sample properties, respectively. As it has been discussed in several previous studies\cite{ashkenasi_1999,sun_2015}, $\Phi_{th}$ can be described by an exponential function
\begin{equation}
\Phi_{th}=\Phi_{\infty}+(\Phi_{1}-\Phi_{\infty})\,e^{-k(\mathbf{N}-1)}%
\label{eq:01},
\end{equation}
where $\Phi_{1}$ is the threshold fluence for a single shot and $\Phi_{\infty}$ is the threshold fluence for an infinite number of pulses. For determining these parameters, we investigated the ablation threshold values for different numbers of shots $\mathbf{N}$, varying the scan velocity for it. In all experiments, the beam focus was positioned on the sample surface. Making several experiments, no significant damage differences ($\backsim$4\%) have been observed from 300 to 500 pulses. Therefore, we can conclude that $\Phi_{\infty}$ converges to $\Phi_{th}(500)$. The experimental values obtained for $\Phi_{1}$ and $\Phi_{\infty}$ are 6.1~J/cm$^{2}$ and 2.9~J/cm$^{2}$, respectively. These fluences are similar to those published by Zhang et al.\cite{Zhang_2016}. Hence, an average incubation factor of $k$~=~0.015 was fitted from experiments. Fig.~\ref{fig:01} shows the ablation threshold fluence as a function of the number of incident pulses for different types of thin film planar waveguides. For the purpose of comparing only the incubation effects of different types of samples, non-dimensional functions of the ablation threshold fluences $\Phi^*_{th}$ are plotted. This expression is shown in Fig.~\ref{fig:01}. To achieve smooth structures, the ablation conditions should be constant, namely, $\partial\Phi^*_{th} / \partial\mathbf{N} \approx 0$. As can be seen in Fig.~\ref{fig:01}, to processing in a \textit{stable ablation regime} samples of PLZT thin films, the number of incident pulses will have to be higher than 325 pulses (at 99\% criterion).

However, in our analysis is more convenient represent the ablation threshold by the kinematic parameters of micromachining. $\mathbf{N}$ can be written as a function of the repetition rate $f$, scan velocity $v$ and spot size $w_0$. Assuming that the focal position and the scan velocity remain constant throughout the process, the number of incident pulses is given by
\begin{figure}[b]
\includegraphics[width=0.45\textwidth]{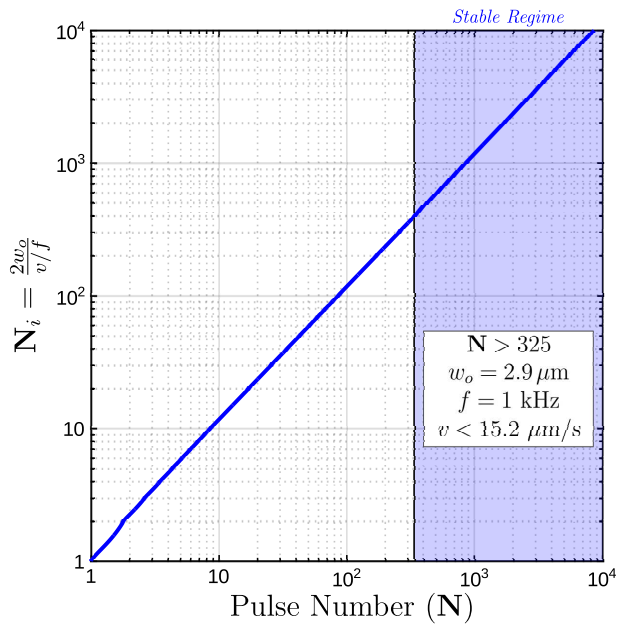}
\caption{\label{fig:02} $\mathbf{N}_i$ as a function of $\mathbf{N}$, see Eq.~(\ref{eq:02}). The stable regime of threshold ablation is shown on the shaded region ($\mathbf{N}>325$).}
\end{figure}
\begin{equation}
\mathbf{N}-1 = 
\begin{cases}
\sum_{i=1}^{\lfloor \mathbf{N}_i \rfloor}\frac{A_o(i \, \mathbf{N}_i)}{\pi w_o^2} &, \text{if \; $\mathbf{N}_i>1$}.\\
0 &, \text{if \; $\mathbf{N}_i \leq1 $}.
\end{cases}
\label{eq:02},
\end{equation}
where 
\begin{equation}
\mathbf{N}_i = \frac{2 w_o}{v/f}
\label{eq:03},
\end{equation}
where $A_o$ is the overlapping area\cite{peyton_2018} and $\mathbf{N}_i$ is the relation between the spot size and the distance among two consecutive pulses ($v/f$). The ablation threshold fluence depends directly on the micromachining kinematic parameters, because it is a function of $\mathbf{N}_i$ as is exhibited in Eq.~(\ref{eq:02}). Therefore, the scan limit velocity for operating in a stable regime of threshold ablation can be calculated from Eq.~(\ref{eq:02}) and (\ref{eq:03}). Fig.~\ref{fig:02} illustrates $\mathbf{N}_i$ as a function of $\mathbf{N}$, here a linear behavior is observed and likewise before, to keep the process in a stable condition is necessary that $\mathbf{N}_i$ be constant. For instance, if the spot size decreases for some reason, then, the distance $v/f$ also must be reduced in the same proportion to hold $\mathbf{N}_i$ constant. This can be achieved either increase the repetition rate (in our case, the laser source limit is 1~kHz) or reduced the scan velocity, which it sometimes decelerates the machining time and turns non-viable the fabrication technique. Obviously, there is a trade-off decision between keeping the stability and the machining speed. The machining kinematic parameters to work in stable regime of threshold ablation is shown in the box of Fig.~\ref{fig:02}. Thus, the scan velocity must be lower than 15.2~$\mu$m/s, given that $f_{\textsc{MAX}}$~=~1 kHz, $w_0$~=~2.9~$\mu$m and $\mathbf{N}>325$.
\begin{figure}[t]
\includegraphics[width=0.48\textwidth]{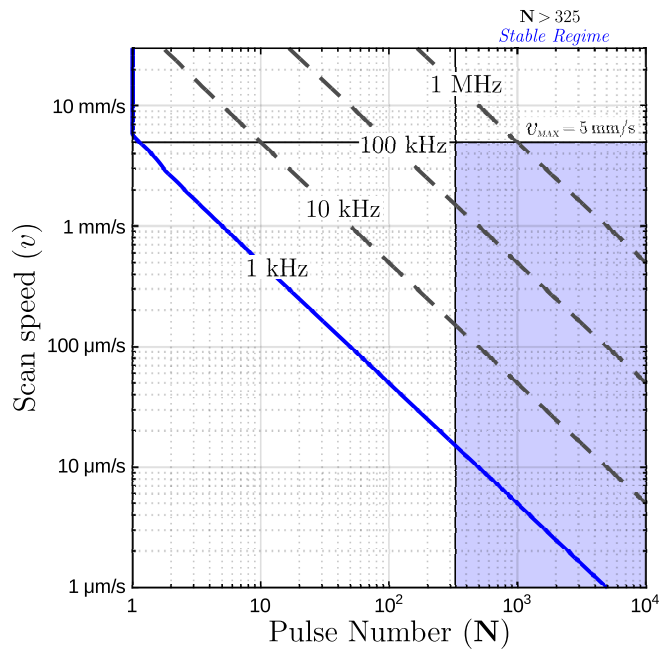}
\caption{\label{fig:03} Number of pulses versus scan speed for $w_0$~=~2.9 $\mu$m. The stable regime of threshold ablation is shown on the shaded region ($\mathbf{N}>325$). Additionally, scan speed as a function of the number of laser shots for different repetition rate are shown.}
\end{figure}

From an engineering point of view, the productive capacity of micromachining technique depends on the system characteristics. Especially, this is affected by kinematic limitations, so it is important to associate the machining parameters with the process stability. The main disadvantage of the femtosecond laser micromachining is the low production capacity since it is directly related to $\mathbf{N}_i$ relation. The distance among two consecutive pulses is limited by the pulse repetition rate, associated to the laser source, and the scan velocity, referred to the motorized station system, and one commonly prevails over the other. Besides, the spot size is a parameter that must be chosen according to the grooves accuracy, pulse energy limitation and sidewall roughness desired. As we can see, it is not an easy task because many factors are involved. If we consider the spot size as a fixed parameter of the system, then the laser and motorized station system should be appropriately chosen to optimize the performance. Indeed, it is desirable a high repetition rate for the laser source; thereby the scan velocity can be increased without losing the stability. Number of pulses versus scan speed is shown in Fig.~\ref{fig:03} for $w_0$~=~2.9 $\mu$m. It can be appreciated in this figure that the scan velocity grows proportionally to the repetition rate. In our case the limitation is given by the maximum repetition rate allowed of the laser system $f_{\textsc{MAX}}$~=~1 kHz. As discussed above, to fabricate ridge optical waveguides on PLZT thin film, the velocity must be lower than 15.2~$\mu$m/s. Nevertheless the motorized station system can reach velocities of up to 5~mm/s, which shows that we are losing productive capacity. In order to improve the machining performance, it is most convenient to match both technological limitations. If the repetition rate would have been 1~MHz, we always can operate in a stable ablation regimen as it was described in Fig.~\ref{fig:03}. In fact the machining time would reduce thousand times. Currently, there are some commercial femtosecond laser system with these characteristics and additionally, up to 1~MHz the thermal diffusion effects can be neglected\cite{Osellame_2012}.

\subsection{Ablation morphology}
\begin{figure}[b]
\includegraphics[width=0.35\textwidth]{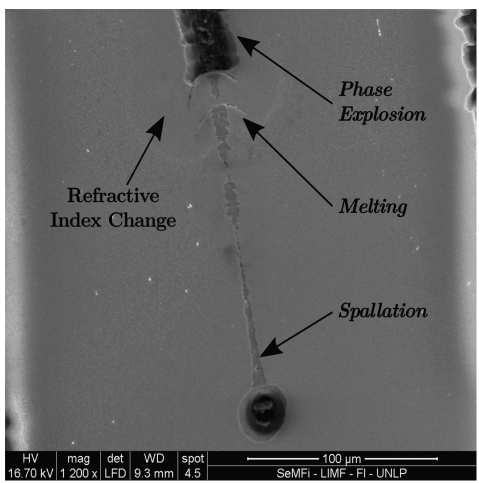}
\caption{\label{fig:04} SEM top picture of the sample ablation. The image corresponds to a machining where the pulse energy is constant at 2.5~$\mu$J and the scan velocity and focus position are changing. The three different regimens are shown.}
\end{figure}

The scan speed and a study of the ablation threshold were determined in the previous section, but still we do not know how much higher the incident fluence should be respected to the threshold fluence. Hence, it is convenient to analyze the dynamical of the process. The ultrafast ablation is classified into three different regimens (\textit{melting}, \textit{spallation} and \textit{phase explosion}), each of them has a probability of occurrence and a particular modification. As expected, the dominant ablation regime depends on the relationship between the incident fluence and threshold fluence. It is defined as the fluence ratio $\mathcal{R}$
\begin{equation}
\mathcal{R}=\Phi/\Phi_{th}
\label{eq:04},
\end{equation}
where $\Phi$ is the incident fluence at the surface. Physically, the regimens are three different hydrodynamics process those relate to the dynamics of phase transitions due to the laser-matter interaction. Below are explained some important features of these regimens: (a) in \textit{melting} there is only surface damage; (b) in \textit{spallation} is obtained smaller craters being a more stable ablation and this process is more controllable than \textit{phase explosion}; (c) where damage in the surroundings of the irradiated area (see Fig.~\ref{fig:04}) is promoted to much higher fluences than the threshold fluences. 

On the other hand, the fabrication of the ridge waveguides is based on removing material of the sample surface. It is desirable to control over the groove dimension and the sidewall roughness, after that the ablation process must be stable. Consequently, the ridge waveguides are fabricated under \textit{spallation} ultrafast ablation regime. Fig.~\ref{fig:04} shows an image taken by microscopy SEM of the sample surface in which can be seen the morphology of the modifications induced. The image corresponds to a machining where the pulse energy is constant at 2.5~$\mu$J, while the scan velocity and focus position are changing with the path. This allows us to modify $\Phi_{th}$ and $\Phi$ at the same time. Fig.~\ref{fig:04} demonstrates that the process can be very unstable to $\mathcal{R}$ changes during the machining.
\begin{figure}[b]
\includegraphics[width=0.48\textwidth]{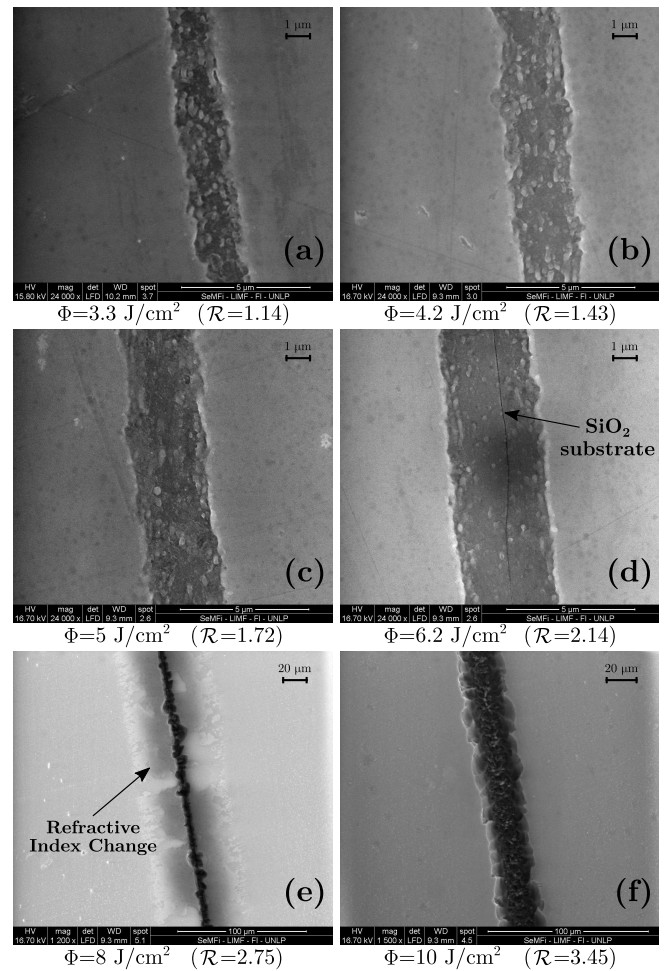}
\caption{\label{fig:05}SEM top picture of grooves ablated at 10~$\mu$m/s ($\Phi_{th}$~=~2.9 J/cm$^{2}$), with the beam waist on the sample surface and at different incident fluences $\Phi$: (a) 3.3~J/cm$^{2}$ (b) 4.2~J/cm$^{2}$ (c) 5~J/cm$^{2}$ (d) 6.2~J/cm$^{2}$ (d) 8~J/cm$^{2}$ and (d) 10~J/cm$^{2}$.}
\end{figure}

Next, we carried out a series of experiments in order to determine the most suitable machining fluence for fabricating waveguides in PLZT/SiO$_2$/Si. Grooves are generated on samples at different incident fluences, varying in each case the laser pulse energy. All tests were performed by scanning at 10~$\mu$m/s and the beam waist positioned on the sample's surface. Under these conditions we are in a steady state fluence regime, where $\Phi_{th}$~=~2.9 J/cm$^{2}$. Six images of grooves ablated at different incident fluences are shown in Fig.~\ref{fig:05}, corresponding to six different fluence ratios. At first glance, it is observed that the width of the groove increases with $\mathcal{R}$, commonly this effect is described by
\begin{equation}
D_c = 2\,w_0\,\sqrt{\frac{1}{2}\,\text{ln}(\mathcal{R})}
\label{eq:05},
\end{equation}
where $D_c $ is the crater diameter. Although this equation is used for static ablation, we will show later that at constant velocities and with a high degree of overlap, equation Eq.~(\ref{eq:05}) quickly approaches to the groove width. Besides, a considerable reduction of the roughness is presented by increasing the incident fluence. This effect is visually appreciable by comparing the sidewall roughness from Fig.~\ref{fig:05} (a) to (d).

In contrast, the depth of the ablated grooves is increased with the incident fluence due to the penetration of the electromagnetic pulse in the sample. Commonly approaches the Beer-Lambert model.
\begin{equation}
H_c = L \, \text{ln}(\mathcal{R})
\label{eq:06},
\end{equation}
where $H_c$ is the crater depth and $L$ is the effective penetration depth of the laser. In this way, we could adjust $\mathcal{R}$ in order to remove the thin film. For example, in Fig.~\ref{fig:05} (d) can be seen a very thin and dark line in the middle of the groove where the ablation reached the silicon dioxide substrate. However, we must be careful, if the ablation threshold fluence of substrate is much lower than the ablation threshold fluence of the film, thus, ablations may occur in the substrate rather than the sample surface. The phase transitions under the film generate a very unstable expansive effect, increasing the roughness of the sidewalls. Since the threshold fluence of SiO$_2$ is noticeably lower than PLZT ($\Phi_{th}$~=~0.9 J/cm$^{2}$ for $\mathbf{N}>30$), it is expected that these effects will occur\cite{rosenfeld_1999,chimier_2011}. Fig.~\ref{fig:05} (e) and (f) show results of several ablations performed at 8 and 10~J/cm$^{2}$, in both the roughness are much greater than with lower energies. Additionally, we have observed damages in the incidence surrounding area, changes in refractive index and a considerable increase of groove width.

The spectrum of the ablation plasma generated in the process was captured in each experiment. Laser-induced breakdown spectroscopy (LIBS) was used as an analysis for contrasted the atomic lines corresponding to silicon. In all cases, atomic lines at different degrees of ionization were observed from the PLZT thin film, and some significant differences according to the incident fluence were found. Since $\Phi$~=~5~J/cm$^{2}$, silicon emissions were identified in the ablation due to the machining depth reached the thickness of the PLZT film. But for fluences greater than 5~J/cm$^{2}$, phase explosions from inside of the sample and an important increase in silicon concentration were observed. Increasing the fluence, we also observe that the concentration of silicon predominates over the PLZT species. After contrasting the results with microscopy SEM, it was observed that the width of the groove and the sidewall roughness come up rapidly with the incident fluence. In conclusion, with the LIBS analysis can be established the best range of fluences to fabrication ridge optical waveguides on thin films of PLZT by femtosecond laser ablation technique. Incident fluences from 3.3 to 6.2~J/cm$^{2}$ and 10~$\mu$m/s of scan speed will be used to operate in stable regime.

\subsection{Roughness study}

In order to achieve smooth structures, a phonomenological study of the roughness is performed. This relates the incident fluence to the roughness considering that the threshold fluence is constant. The latter was supported considering that by using a scan speed at 10 $\mu$m/s, $\Phi_{th}$ goes to $\Phi_{\infty}$. We propose a model for that in which two factors affect the machining quality, one is due to deterministic effect and other due to randomness effect. As the roughness can be understood as a result about changes in the machining track, then it is convenient to define the width $\mathrm{W}$ of ablated groove as
\begin{equation}
\mathrm{W} = \mathrm{\overline{W}}+\delta+\xi
\label{eq:07},
\end{equation}
where $\mathrm{\overline{W}}$ is the mean width of the groove, $\delta$ is a deterministic function of the trajectory and $\xi$ is a random variable. 
\begin{figure}[htb!]
\includegraphics[width=0.4\textwidth]{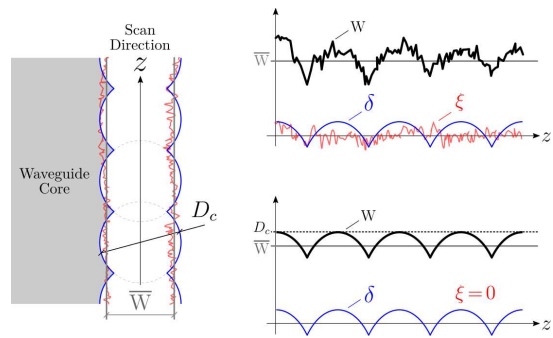}
\caption{\label{fig:06}Schematic representation of the proposed roughness model. On the left is a top view of an ablated groove, which covers a part of the waveguide. With blue lines the deterministic effects and red lines the random effects are represented. The mean width of the groove sketches with gray lines. On the right, each term of Eq.~(\ref{eq:07}) is plotted with and without the stochastic effects.}
\end{figure}

The function $\delta$ is related to discontinuity of laser micromachining technique. Indeed, since the groove is done by pulse overlapping, it could show a periodic peaks in the sidewall of tracks related to the sample displacement\cite{choi_2007}. The random variable $\xi$ is associated with changes in the ablation conditions between one pulse and other. This is connected to laser deviation conditions (incident fluence), the dynamics of the sample interaction (threshold fluence) and the material defects. Namely, $\xi$ has directly correlated with random fluctuations of the fluence ratio $\mathcal{R}$. Even though $\xi$ would not had been removed  completely, later we will show that it can be reduced by a suitable selection of the fluence ratio, through multiple scans. These ideas are sketched in Fig.~\ref{fig:06}, in which random and deterministic effects are drawn in red and blue lines, respectively. Next, we will appreciate that the deterministic effects may be neglected if the pulse number $\mathbf{N}_i$ is large enough and prove that the roughness is mainly stochastic.

If assuming $\xi$=0, we can clear $\delta$ of Eq.~(\ref{eq:07}). Replacing the maximum value of $\mathrm{W}$ for that case then we get  
\begin{equation}
	\delta = \mathrm{W}-\mathrm{\overline{W}} \leq D_c  - \mathrm{\overline{W}}
\label{eq:08}.
\end{equation}
The inequality of Eq.~(\ref{eq:08}) is an upper bound of the $\delta$ function. Next, we define a function of normalized roughness $\delta^\textbf{*}$ as
\begin{equation}
\delta^\textbf{*} = 1-\frac{\mathrm{\overline{W}}}{D_c}
\label{eq:09}.
\end{equation}

Combining Eq.~(\ref{eq:08}) and Eq.~(\ref{eq:09}), the following relation is established
\begin{equation}
\delta \leq D_c \, \delta^\textbf{*}
\label{eq:10}.
\end{equation}

Therefore, if we show that $\delta^\textbf{*}$ tends to zero for a large pulse number, then we can demonstrate that $\delta$ also tends to zero. Since $\mathbf{N}_i$ is larger than 1 and using geometric models\cite{peyton_2018}, the mean width of the groove is given by
\begin{equation}
\mathrm{\overline{W}} = D_c \, \left( \frac{\mathbf{N}_i^{-1}}{\frac{\pi}{2}-\text{arccos} \left( \mathbf{N}_i^{-1} \right)} \right)
\label{eq:11}.
\end{equation}

Replacing Eq.~(\ref{eq:11}) in Eq.~(\ref{eq:09}), it is obtained a relation only of the pulse number
\begin{equation}
\delta^\textbf{*} = 1  - \frac{\mathbf{N}_i^{-1}}{\frac{\pi}{2}-\text{arccos} \left( \mathbf{N}_i^{-1} \right)}
\label{eq:12}.
\end{equation}
\begin{figure}[t]
\includegraphics[width=0.45\textwidth]{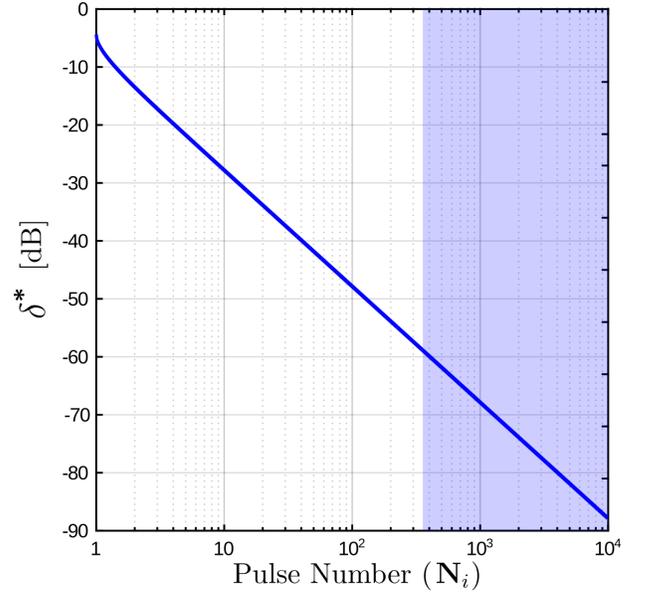}
\caption{\label{fig:07}Normalized roughness $\delta^\textbf{*}$ as a function of $\mathbf{N}_i$, see Eq.~(\ref{eq:12}). The stable regime of threshold ablation is shown on shaded region.}
\end{figure}

Therefore, we prove the origin random of the roughness by taking the limit of $\delta^\textbf{*}$ for $\mathbf{N}_i$ going to infinity
\begin{equation}
\lim_{\mathbf{N}_i \to \infty} \delta = \lim_{\mathbf{N}_i \to \infty} \delta^\textbf{*} = 0 
\label{eq:13}.
\end{equation}

The normalized roughness $\delta^\textbf{*}$ as a function of $\mathbf{N}_i$ is evaluated in Fig.~\ref{fig:07}, in which the roughness decreases at 20~dB per decade. For example, $\mathbf{N}_i$ must be higher than 300 to stabilize the threshold fluence as we showed previously. The normalized roughness for that value is lower than -60~dB regarding the crater diameter, see Eq.~(\ref{eq:10}). Then the roughness $\delta$ will be in the picometers order if we assume that $D_c$ is in the microns order. Under this condition, the deterministic effects can be neglected.
\begin{figure}[b]
\includegraphics[width=0.43\textwidth]{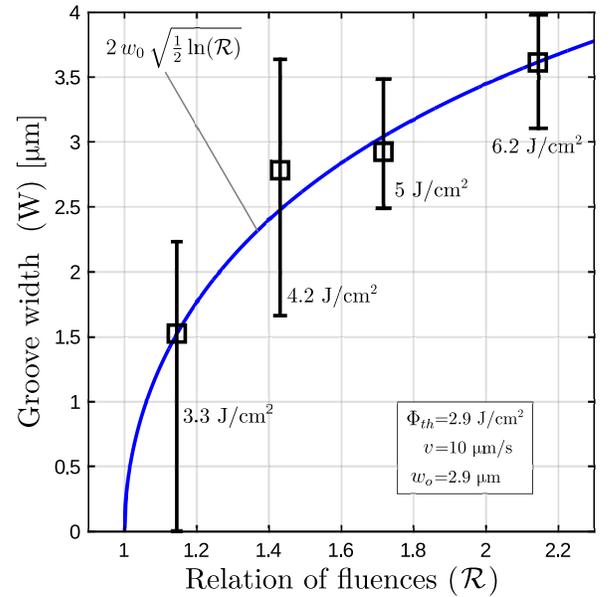}
\caption{\label{fig:08}Experimental data (black lines) and theoretical correlation (blue line) of the groove width. Incident fluences from 3.3 to 6.2~J/cm$^{2}$ and a scan speed of 10~$\mu$m/s were used to operate in the stable regime.}
\end{figure}

Otherwise, the roughness of ablated grooves recorded at 10~$\mu$m/s and by using fluences from 3.3 to 6.2~J/cm$^{2}$ have been measured by SEM microscopy. Fig.~\ref{fig:08} shows the experimental results, there is a strong correlation between the average width $\mathrm{W}$ of the grooves and the function of the crater diameter shown in Eq.~(\ref{eq:05}). The roughness is directly related to instantaneous changes in diameter, as black bars shown in Fig.~\ref{fig:08}. Also, we can see that the roughness is inversely proportional to the fluence ratio. Combining our roughness model and the crater diameter correlation, Eq.~(\ref{eq:07}) is reduced to
\begin{equation}
\mathrm{W} = D_c + \xi
\label{eq:14}.
\end{equation}

The roughness typically is related to scattering losses by the mean square error of the waveguide. It can be obtained from measurements as $\xi_\textsc{rms}$ and it is given by\cite{marcuse_1969,ladouceur_1994}
\begin{equation}
\xi_\textsc{rms} = (\mathrm{W} - \mathrm{\overline{W}})_\textsc{rms}
\label{eq:15}.
\end{equation}

On the other hand, the fluence ratio is the main source of roughness due to the random origin of the laser-matter interaction process. Therefore, we could explain the roughness of micromachining as a function of the mean square error of the fluence ratio $\mathcal{R}_\xi$. Using a first-order approximation of the crater diameter and $\mathcal{R}_\xi$, the following expression is proposed
\begin{equation}
\xi_\textsc{rms} \approx \mathcal{R}_\xi \, \dfrac{\partial D_c}{\partial \mathcal{R}}
\label{eq:16},
\end{equation}

where $\mathcal{R}_\xi$ is an \textit{ablation quality factor}. See that, the approximation is valid for small values of $\mathcal{R}_\xi$. Deriving Eq.~(\ref{eq:05}) with respect to $\mathcal{R}$ and replacing it in Eq.~(\ref{eq:16}), the roughness is finally obtained
\begin{equation}
\xi_\textsc{rms} \approx \mathcal{R}_\xi \, \dfrac{w_0}{\mathcal{R}\, \sqrt{2 \, \text{ln}(\mathcal{R}) }}
\label{eq:17},
\end{equation}
where $\mathcal{R}_\xi$ must be adjusted. 
\begin{figure}[htb!]
\includegraphics[width=0.43\textwidth]{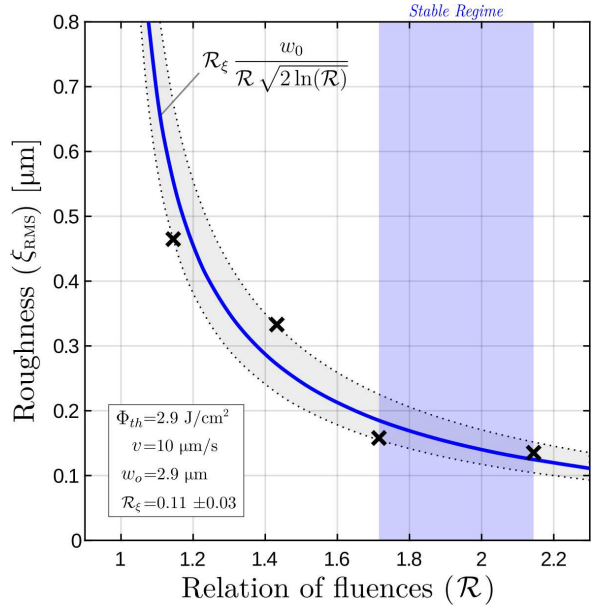}
\caption{\label{fig:09}Experimental data (black cross) and theoretical correlation (blue line) of the groove width. Incident fluences from 3.3 to 6.2~J/cm$^{2}$ and a scan speed of 10~$\mu$m/s were used. Stable regime of ablation is shown on the shaded region.}
\end{figure}

Indeed, the ablation quality factor was fitted with the measurements. Fig.~\ref{fig:07} shows the experimental results and theoretical correlation. The ablation quality factor $\mathcal{R}_\xi$ is $0.11\pm0.03$. From this analysis, we have set the most suitable fluence range for achieving smooth structures. As it was done in the threshold ablation section, here we were determined that the fluence ratio range from 1.72 to 2.14 is a stable ablation regime (see shaded area in Fig.~\ref{fig:08}). In summary, we first have optimized the scan velocity and later the incident fluence for the femtosecond laser ablation. Finally, ridge waveguides on PLZT thin films were fabricated at 10~$\mu$/s and $\Phi$~=~5~J/cm$^{2}$ ($\mathcal{R}$~=~1.72) to avoid occasional explosions due to changes in substrate phase.

\subsection{Waveguides characterization}

Once defined the most suitable micromachining parameters, we proceed to fabricate the ridge optical waveguides and measure their performance. In order to characterize the quality of machining, the coupled modes and scattering losses were analyzed. The same experiment proceeding have been repeated. The beam focus was placed on the sample surface and a track distance of 20~$\mu$m was used. An effective width of waveguides around 17~$\mu$m was measured. Under these conditions and using a laser with 532~nm of wavelength, the waveguides are multi-mode and support TE/TM modes. Additionally, we have fabricated two types of devices to verify the effect of multiple scan technique, waveguides with a single scan and other with several round trips (multiple scans).
\begin{figure}[b]
\includegraphics[width=0.48\textwidth]{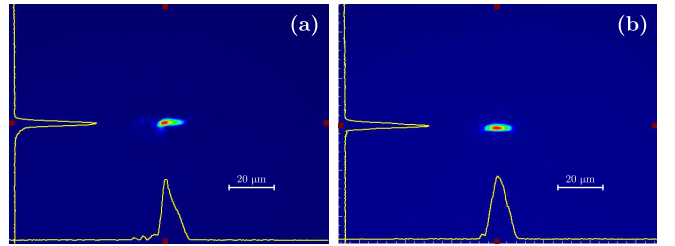}
\caption{\label{fig:10}Measured TE/TM modal profiles of ridge waveguides fabricated by single scan (a) and multiple scans (b). An effective width of waveguides around 17~$\mu$m and a laser with 532~nm of wavelength were used.}
\end{figure}

The mode profile corresponding to ridge waveguides by single scan Fig.~\ref{fig:10} (a) and six scans Fig.~\ref{fig:10} (b) are shown. There is a considerable difference in the distribution of intensity, obviously it indicates that the ablated grooves are morphologically different according to the number of scans. The mode profile of waveguides performed by multiple scans is similar to the theoretical mode profile of strip waveguides, while it is not achieved with a single scanning. Therefore the ridge waveguides made by multiple scans exhibit better geometric characteristics. These are interesting for photonic applications due to the performance of waveguides can be precisely predicted and the decoupling losses will be lower. There are differences because, when the ablation process is repeated many times, the pulse penetration and the generation of groove are saturated, and hence the width and the depth randomness of grooves tend to cancel. However, more debris was observed on the surface of the samples which contributes to the surface scattering losses. The sidewall of grooves ablated by multiple scans is much more vertical than grooves fabricated by a simple scan.
\begin{figure}[t]
\includegraphics[width=0.35\textwidth]{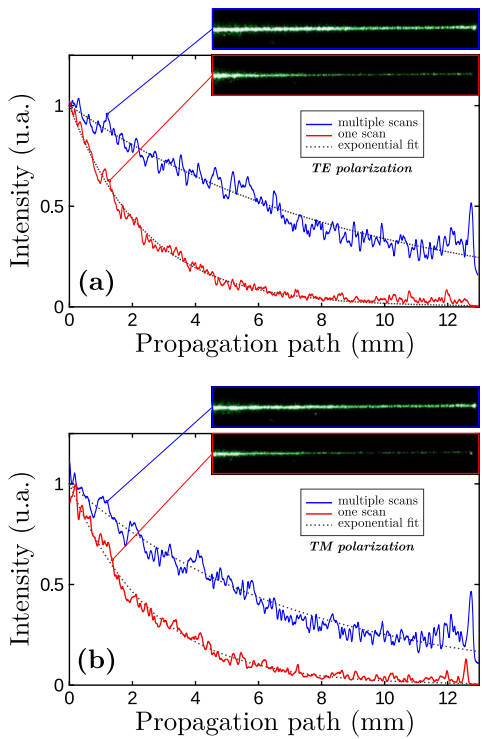}
\caption{\label{fig:11}Scattering losses at 532~nm from a ridge waveguide with 17~$\mu$m and 12~mm length fabricated by femtosecond laser ablation, translation speed of 10~$\mu$m/s, and fluence $\Phi$~=~5~J/cm$^{2}$. Single (red) and multiple scan (blue) for TE (Fig.~a) and TM (Fig.~b) laser input were used.}
\end{figure} 
                     
Referred to the roughness the waveguides fabricated by multiple scans also reported better capabilities than those made with a single scan. The scattering losses of the ridge waveguides are shown in Fig.~\ref{fig:11}. They can be used to relate the roughness. The scattering images were captured using a reflection digital microscope positioned over the samples. By a theoretical analysis of scattering losses\cite{payne_1994}, which relates scattering effects and surface roughness, we could be determined a significant reduction in random effects due to the use of multiple scans.The scattering loss of waveguides can be converted to an absorption coefficient by fitting two relative exponential functions, like as shown in Fig.~\ref{fig:11}. From this characterization we have calculated that the scattering losses for multiple scans are reduced in 2.5~dB for TE modes and 1.9~dB for TM modes. This is also related to the average of the random effects of ablation, since to when multiple scans are performed, the average size of grooves tends to the maximum deviation of roughness. However, if multiple scans are used then the machining process time is considerably increased being proportional to the number of scans. In our case, each scan took 45 minutes, so for six scans was delayed 4 hours and 30 minutes approximately, and therefore this difference makes that the technique for reducing the roughness is not the most appropriate resource. Currently, the optical fiber fs system with MHz range of repetition rate is commercially produced, so by using this kind of equipment the micromachining procedure presented in this paper can be performed by short time consuming. Under this situation, the method shown here can be very competitive with standard ones in order to define optical circuits on thin films considering mass production.

\section{Conclusions}
In conclusion, we have reported, for the first time of our knowledge, on the micromachining of ridge optical waveguides on PLZT thin film by femtosecond laser ablation. We also have studied the relationship between machining parameters and morphology of ablated grooves with the goal of reducing the roughness of sidewall. In this sense, we have investigated firstly the incubation effect on the threshold fluence and its connection with the kinematic of the ultrafast ablation process. At low velocities, the threshold fluence is established in a regime stable and smooth structures are more likely promoted. The threshold fluence converges to a common value of $\backsim$2.9~J/cm$^{2}$ for velocities lower than 15.2~$\mu$m/s, in particular, we have used 10~$\mu$m/s. We have analyzed subsequently the morphology of ablation regimens for different incident fluences, using SEM microscopy and laser-induced breakdown spectroscopy, in order to establish the suitable fluence range for photonics applications. Thus a phenomenological model of roughness has been proposed from which the best result was reported at 5~J/cm$^{2}$. This model is innovative from several points of view, nevertheless the most important is the correlation exposed between roughness and incident fluence through an \textit{ablation quality factor}, see Eq.~(\ref{eq:17}). The ridge optical waveguides have been carried out by a simple and multiple scan with the purpose of characterization their performance. The waveguides made with six scans have achieved cut down the scattering losses in 2.5~dB for TE modes and 1.9~dB for TM modes and additionally shows an improvement in their geometrical morphology. In summary, this work explains how to define the key micromachining parameters for achieving smooth structures, what must be considered and which is the link among roughness and random process inherent to femtosencond laser ablation.

\begin{acknowledgments}
This work was partially supported by the Agencia Nacional de Promoci\'on Cient\'ifica y Tecnol\'ogica (Argentina) under the project PICT-2016-4086 and by Universidad Nacional de Quilmes (Argentina) under project PPROF-901-2018. The authors also would like to thank to Dra. Nora Pellegri from Instituto de F\'isica de Rosario (IFIR) for supplying the samples used in this work.
\end{acknowledgments}

\bibliography{Phenomenological_studies_of_femtosecond_laser_ablation_on_optical_thin_films_for_integrated_photonics}

\end{document}